\documentclass[12pt,preprint]{aastex}

\def\la{\mathrel{\hbox{\rlap{\hbox{\lower4pt\hbox{$\sim$}}}\hbox{$<$}}}}
\def\ga{\mathrel{\hbox{\rlap{\hbox{\lower4pt\hbox{$\sim$}}}\hbox{$>$}}}}

\newcommand{\be}{\begin{eqnarray}}
\newcommand{\ee}{\end{eqnarray}}

\def\nuc#1#2{\relax\ifmmode{}^{#1}{\protect\mbox{#2}}\else${}^{#1}$#2\fi}
\newcommand{\etal}{et al.}

\newcommand{\kmps}{\ensuremath{\mbox{km}~\mbox{s}^{-1}}}

\newcommand{\msol}{\ensuremath{{\mbox{M}_\odot}}}

\newcommand{\nni}{\nuc{56}\mbox{Ni}}

\def\ang{\mbox{\AA}}
\def\Tmod{\ensuremath{T_{\mbox{model}}}}

\def\tstd{\ensuremath{\tau_{\mbox{std}}}}

\newcommand{\phx}{\texttt{PHOENIX}}

\newcommand{\gamray}{$\gamma$-ray}

\newcommand{\SiII}{Si~II}
\newcommand{\SiIIred}{\SiII$\lambda 6355$}
\newcommand{\SiIIblue}{\SiII$\lambda 5972$}

\received{} 
\accepted{} 
\journalid{}{} 
\articleid{}{} 
\shortauthors{Baron, E. et~al.}
\shorttitle{Spectral Modeling of SNe Ia Hydro Models}

\begin{document}

\title{Spectral Modeling of SNe Ia Near Maximum Light: Probing 
  the Characteristics of Hydro Models}

\author{ E.~Baron,\altaffilmark{1,2}\email{baron@nhn.ou.edu}
  Sebastien Bongard,\altaffilmark{1,3} David
Branch,\altaffilmark{1}\email{branch@nhn.ou.edu}  and  Peter
H.~Hauschildt\altaffilmark{4}\email{yeti@hs.uni-hamburg.de}
 }

\altaffiltext{1}{Homer L.~Dodge Department of Physics and Astronomy,
  University of 
Oklahoma, 440 West Brooks, Rm.~100, Norman, OK 73019-2061, USA}

\altaffiltext{2}{Computational Research Division, Lawrence Berkeley
  National Laboratory, MS 50F-1650, 1 Cyclotron Rd, Berkeley, CA
  94720-8139 USA}

\altaffiltext{3}{Institute de Physique Nucl\'eaire Lyon, B\^atiment Paul Dirac
Universit\'e Claude Bernard Lyon-1
Domaine scientifique de la Doua
4, rue Enrico Fermi
69622 Villeurbanne cedex, France}

\altaffiltext{4}{Hamburger Sternwarte, Gojenbergsweg 112,
21029 Hamburg, Germany}


\begin{abstract}
  We have performed detailed NLTE spectral synthesis modeling of 2
  types of 1-D hydro models: the very highly parameterized
  deflagration model W7, and two delayed detonation models. We find
  that overall both models do about equally well at fitting well
  observed SNe~Ia near to maximum light. However, the Si~II 6150
  feature of W7 is systematically too fast, whereas for the delayed
  detonation models it is also somewhat too fast, but significantly
  better than that of W7. We find that a parameterized mixed model
  does the best job of reproducing the Si~II 6150 line near maximum light and
  we study the differences in the models that lead to better fits to
  normal SNe Ia. We discuss 
  what is required of a hydro model to fit the
  spectra of observed SNe Ia near maximum light.
\end{abstract}


\section{Introduction}
Intense astronomical interest has been focused on Type Ia supernovae
since it was recognized long ago \citep{wilson39,kowal68} that they
are good ``standard candles'' and hence are useful cosmological
probes. The reliability of SNe Ia as distance indicators improved
significantly with the realization that the luminosity at peak was
correlated with the width of the light curve \citep{philm15} and hence
that SNe~Ia are correctable candles in much the same way that Cepheids
are \citep{philetal99,goldhetal01,rpk95}.  With the discovery of the
``dark energy'' this interest has been further piqued
\citep{riess_scoop98,garnetal98,perletal99}. Nevertheless, we still
don't know what the stellar progenitors for SNe Ia are, nor how they
evolve with redshift. In addition, hydrodynamical modeling of SNe Ia
explosions has now entered the sophisticated realm of fully 3-D hydro
models \citep{RH05a,gko04a,gko05}. In spite of the sophistication of
the modeling, the enormous dynamic range involved in modeling the
fully turbulent flame propagation still requires the use of sub-grid
models, or direct numerical simulation of small portions of the star
\citep{zingale05,bellflame04a,bellflame04b}. In addition the 3-D
models produce results that at first sight appear to be worse
representations of the true phenomenon than the highly parameterized
1-D models. Deflagration models in particular appear to lead to ejecta
configurations where unburnt C+O is mixed with burned material in such
a way that the spectra produced by these models would differ
significantly with observations. \citet{kozma05} showed that late-time
spectra of a particular 3-D deflagration model would lead to strong
[C~I] and [O~I] lines about 300 days after explosion that are
unobserved in SNe~Ia. The propensity for 3-D deflagration models to
leave unburned C+O in the center has led to the suggestion that a
detonation to deflagration transition \citep{gko04a,gko05} or a
confined detonation \citep{PCL04} is required to make the 3-D
hydrodynamical models look more like the 1-D models which have
stratified compositions. 

Clearly the goal of SNe~Ia explosion modeling is to perform full
3-D radiation hydrodynamics simulations with full reaction
networks. Given present computer resources such realistic calculations
are not yet feasible. Some parameterized 3-D synthetic spectral
calculations have been performed
\citep{KP05,kasen01el03,kasen_hole03,thomas02}, but calculations
involving full detailed line and continuum NLTE still can only be
performed in spherical symmetry.

\section{Motivation}

In previous work we have studied the detailed NLTE synthetic spectra
of the parameterized 1-D model W7 \citep{nomw7,nomw72} as well as the
synthetic spectra 
produced by 1-D delayed detonation models \citep{hwt98,iwamoto99} using the
detailed generalized stellar atmosphere code \phx\ 
\citep{l84a01,l94d01,nughydro97,nug1a95}. Since that work, several
improvements have been made to the \phx\ code and more 1-D delayed
detonation models have been calculated. In particular, more species
have been added into NLTE in \phx, and the treatment of the large
dynamic range required to treat multiple ionization stages in NLTE has
been improved through the use of up to 256-bit
precision in solving the linear system for the rate equations, and in
the EOS, making use of the QD package \citep*{QD01}. Thus, 
up-to-date calculations should provide a better representation of the
actual synthetic spectrum predicted by a particular hydrodynamical model.
The hydrodynamic model we use is a somewhat modified version of the
model presented in \citep{branch81b85} in that we have extended the
C+O layer from 22,000~\kmps to 30,000~\kmps using a density law $\rho
\propto e^{-v/v_e}$ with $v_e = 2700$~\kmps. The extension is
necessary in order to for the model to be optically thin in the UV and
for it to reproduce the high velocity features that have been
found in numerous SNe~Ia \citep[for example][]{hatano94D99}. This
extension does 
not affect the total mass of the model.

Consequently, it makes sense to reexamine the synthetic spectra of both W7 and
modern delayed detonation models. Here we focus on two particular
delayed detonation models of \citet{HGFS99by02} 5p0z22.25 which has
$\Delta m_{15} = 1.00$ and 5p0z22.16 which has $\Delta m_{15} = 1.26$
\citep[see Table 2 of][]{HGFS99by02}. 

\citet{nug1a95} modeled a homogeneous model with a similar density
structure to that of W7, and compositions obtained by mixing those of
W7 above 8,000~\kmps\ using an older
version of \phx. They found that this model did a reasonably good job of
reproducing the +5 day spectrum of SN~1992A and the maximum light
spectrum of SN~1981B, but they treated only Ca~II, Mg~II, and Na~I in NLTE.

\citet{l94d01} used the fully stratified W7 hydro model and improved
gamma-ray deposition to model a time series of spectra of
SN~1994D. the species treated in NLTE were: H~I, He~I-II, C~I, O~I, Ne~I, Na~I,
Mg~II, Si~II, S~II, Ti~II, Fe~II, and Co~II.  They also used an earlier
version of \phx\ which did not handle the dynamic range associated
with multiple ionization stages in NLTE in as sophisticated a manner
as does the current version.  In that work they
were able to obtain acceptable fits to SN~1994D with W7, up until
about March 18 (-3 days). After that epoch the quality of the fits
deteriorated significantly, and prior to March 15, 1994, the Ca H+K
feature was poorly fit by W7.

We have chosen to model the two supernovae that we had previously
modeled, SN~1992A and SN~1994D, and the very  well-observed
SN~1999ee. All of these supernovae were well observed, SN~1992A has a
\emph{HST} UV+optical spectrum and thus offers wide wavelength
coverage. The basic properties of these supernovae are listed in
Table~\ref{tab:snparams}. While both SN~1992A and SN~1994D were
moderately fast decliners, SN~1999ee was somewhat slower. The
$B_{\mathrm{max}}-V_{\mathrm{max}}$ colors for all these supernovae is
close to zero, but D.~Branch et al.~(in preparation) find that none of
them is directly related to each other when the widths of the
\SiIIblue\ and \SiIIred\ lines are compared. 

\subsection{Methods}

We have used the multi-purpose spectrum synthesis and model atmosphere
code \phx, version \texttt{13.11}\, \citep[see][and references
therein]{hbjcam99,hbmathgesel04}. \phx\ has been designed to
accurately include the 
various effects of special relativity important in rapidly expanding
atmospheres, like supernovae.  Ionization by non-thermal electrons
from $\gamma$-rays from the nuclear decay of \nni\ that powers the
light curves of SNe~Ia is taken into account.  We have used, an
updated method for calculating the $\gamma$-ray deposition using a
solution of the spherically symmetric radiative transfer equation for
$\gamma$-rays with \phx.  We used an effective \gamray\ opacity,
$\kappa_{\gamma} = 0.06$~Y$_e$~cm${^2}$~gm$^{-1}$ \citep{cpk80} for
all calculations. The following species were treated in NLTE: H~I,
He~I-II, C~I-II, O~I-III, Ne~I, Na~I-II, Mg~I-III, Si~I-III, S~I-III,
Ca~II, Ti~II, Fe~I-III, and Co~II. The atomic data used to construct
the model atoms is taken from a variety of sources. The energy levels
and radiative bound-bound cross sections are obtained from the work of
Kurucz \citep{kurucz93,cdrom22,cdrom23}, the bound-free rates are based on data from the Opacity
Project, the Iron Project, and other sources. The collisional rates
are based on the Opacity project, \citet{reilman79}, and other
sources.

The model structure for both W7 and the delayed-detonation models was
taken from the output of the hydrodynamical models after they had
reached the homologous phase. The abundances were also taken directly
from the hydro models. The models were expanded homologously, and
the decay of the radioactive species accounted for. 

A rise time of
20 days after explosion to maximum light in $B$
\citep[e.g.,][]{riessetalIart00,akn00} was assumed for all
calculations. For W7, the hydrodynamic output was 
extended from $\sim 24000$~\kmps\ to 30000~\kmps\ with the unburned C+O
white dwarf composition as in previous \phx\ calculations using W7
\citep{nughydro97,nugphd,lentzmet00,l94d01}.

At each epoch, we have adjusted the total bolometric luminosity in the
observer's frame and we present results which provide the best fit
(chi-by-eye) to match the shape and color of the observations. In each
model the radiative transfer, the energy balance and NLTE rate
equations are fully converged. About 150 models were constructed for
the work reported here. Our methods involve the full solution to the
NLTE radiative transfer problem including energy
conservation. Gamma-ray deposition is treated as an explicit source
term in the generalized equation of radiative equilibrium and no lightbulb is
assumed. 

In selecting the best fit, we first strive to fit the overall shape
(or colors) and then we strive to fit the lineshape of selected lines.
We are developing statistical tests to improve the sensitivity and we
note that we are sensitive to much smaller variations of order 0.02
mag when this method is applied to SNe~II
\citep{mitchetal87a02,mitchetal87a01,bsn99em04}.

\begin{deluxetable}{lcc}
\tablecolumns{3}
\tablewidth{0pc}
\tablecaption{\label{tab:snparams}Parameters for Modeled SNe}
\tablehead{\colhead{SN} &
\colhead{$\Delta m_{15}$}    &   \colhead{$B_{\mathrm{max}}-V_{\mathrm{max}}$}}
\startdata
SN 1992A&1.47&$0.02$\\
SN 1994D&1.31&$-0.05$\\
SN 1999ee&0.94&$-0.02$\\
\enddata
\tablenotetext{a}{The values for SN 1992A and SN 1994D were taken from
  \citet{DR94D99} and for SN 1999ee from \citet{stritz99ee99ex02}.}
\end{deluxetable}

\section{Results}

\subsection{SN 1994D: Maximum Light}

SN 1994D, in NGC 4526, was discovered 2 weeks before maximum
brightness \citep{iauc94d}.  It was one of the best observed SNe~Ia,
with near-daily spectra starting 12 days before maximum brightness
(-12 days) and continuing throughout the photospheric phase.  SN~1994D
has been well observed photometrically
\citep{richsn94d,patat94D96,meikle94d91t,tsv94d94i} and
spectroscopically
\citep{filipasi97,patat94D96,meikle94d91t}. \citet{wang97} found no
significant polarization in SN~1994D 10 days before maximum light.
\citet{cum94d96} placed a limit on a solar-composition progenitor wind
of $1.5 \times 10^{-5}$~M$_{\odot}$~yr$^{-1}$ for a 10 \kmps\ wind.
SN~1994D has been previously modeled with synthetic spectra and light
curves by several groups
\citep{hatano94D99,hofsn94d,meikle94d91t,mazz94d,bran94d05}.

\begin{figure}
\includegraphics[width=.6\textwidth,angle=90]{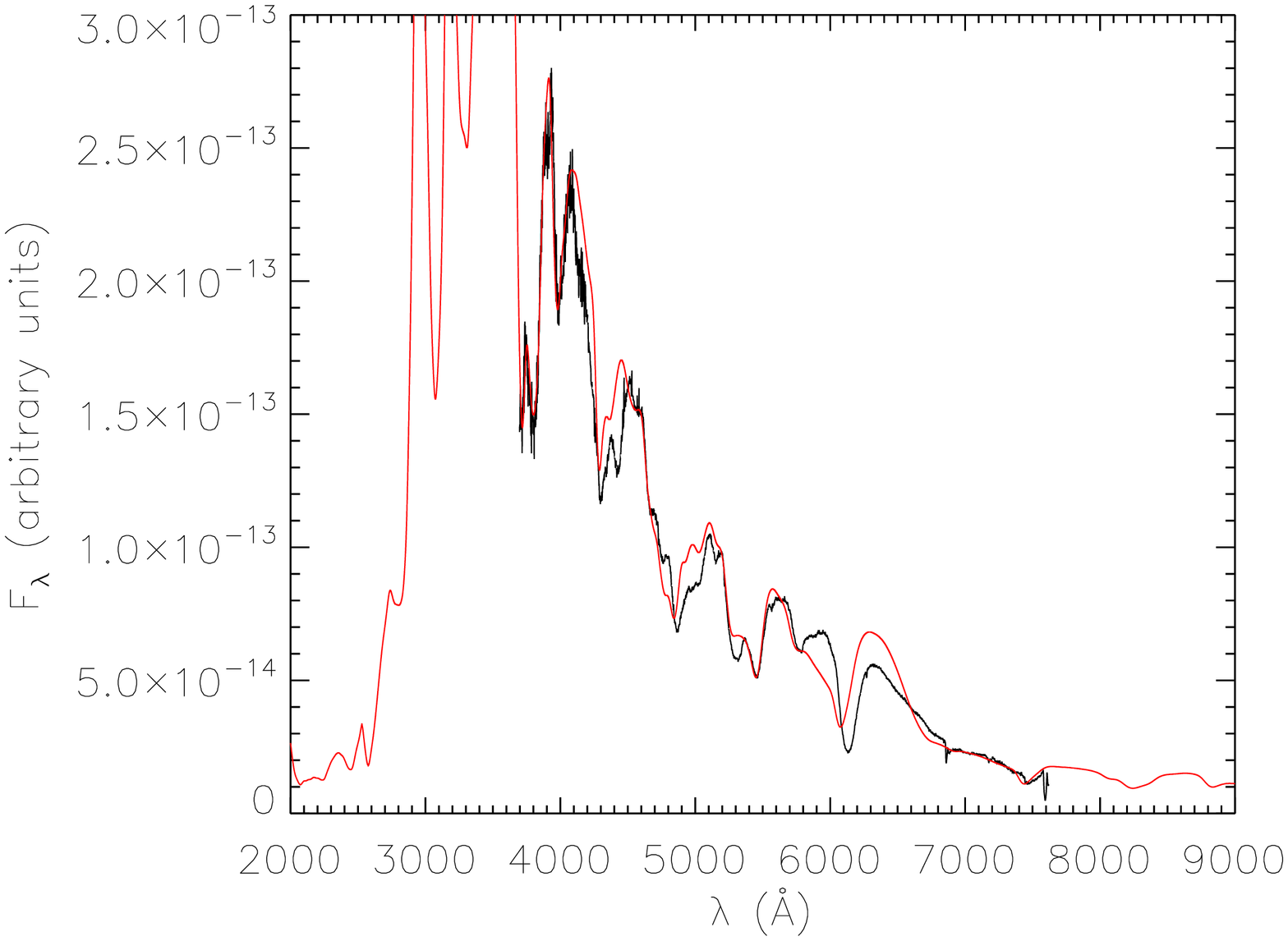}
\caption{\label{fig:w7_94d_mar21}The synthetic spectra of a full NLTE
  model of W7, 20 days after explosion, is compared to the observed
  spectrum of SN~1994D on March 21, 1994 (the time of B maximum). In
  this and subsequent figures on SN~1994D the observed spectra have
  been corrected for redshift assuming a velocity of 448~\kmps\ and a
  reddening of $E(B-V) = 0.06$.}
\end{figure}

Figure~\ref{fig:w7_94d_mar21} displays our best fit for W7 compared to
the spectrum of SN~1994D observed on March 21, 1994 (day -1). The observed
spectra for SN~1994D have been dereddened by $E(B-V) = 0.06$
\citep[cf.][]{DR94D99}, and 
de-redshifted using a recession velocity of 448~\kmps. The fit is not
bad and confirms that W7 is a reasonable first starting point for
comparing hydrodynamic calculations to observations. The overall shape
well reproduces the observed spectrum. The Ca~II H+K line is nicely
fit and the S~II ``W'' is reasonably well fit.  Nevertheless, the fit
is poorest in the region around the defining feature of SNe~Ia, Si~II
$\lambda 6355$. The Si~II line is prominent and strong enough, but the
absorption forms much too far out in the ejecta. The weaker Si~II
$\lambda 5972$ line is barely evident in the synthetic spectrum. This
may be due to the extended blue wing of Si~II $\lambda 6355$, which as
we discuss below is likely due to the interaction of Fe~II/III line
blending and the Si~II line.

\citet{bran94d05} performed a detailed direct comparative analysis of
the spectra of SN~1994D and even with a highly parameterized synthetic
spectra code were not able to produce perfect fits at this epoch. When
referring to features in the observed spectrum we will use the
their line identifications.

In the work of \cite{l94d01} the Si~II $\lambda 5972$ line is
prominent, although it is not strong enough in their model, it is
still stronger than we find, this difference is almost certainly due
to the more complete treatment of ionization stages in this work.

\begin{figure}
\includegraphics[width=.6\textwidth,angle=90]{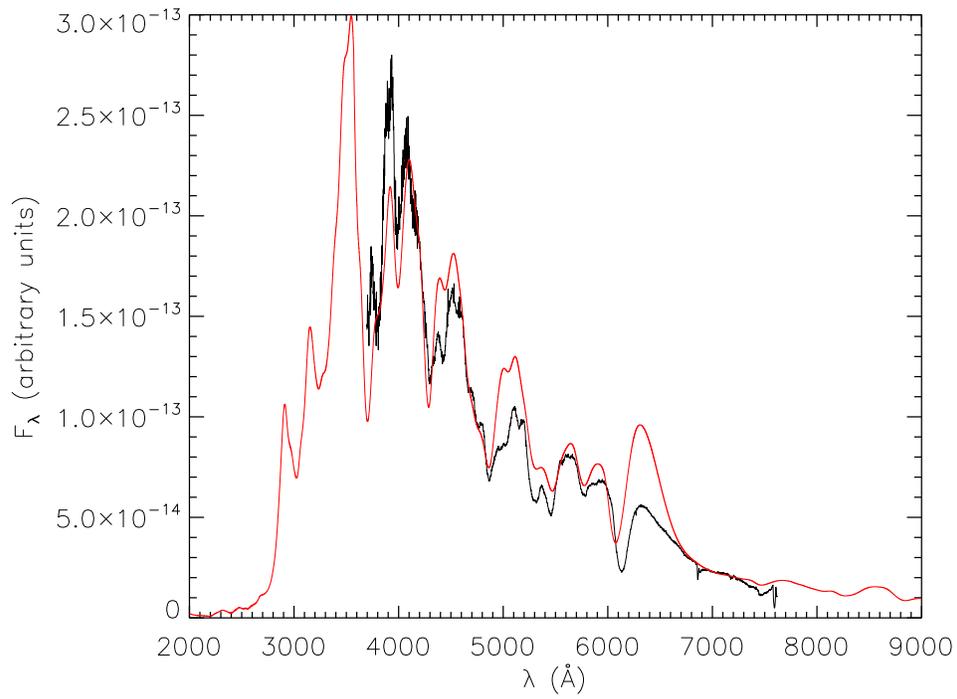}
\caption{\label{fig:pah_94d_mar21}The synthetic spectra of a full NLTE
  model of hydro model 5p0z822\_16 (red line), 20 days after explosion,
  is compared to the observed 
  spectrum of SN~1994D on March 21, 1994 (the time of B maximum).}
\end{figure}

Figure~\ref{fig:pah_94d_mar21} displays the synthetic spectrum of our
best fit to SN 1994D at maximum light using the 5p0z22.16 model of
\citet{HGFS99by02}. The bolometric luminosity is the same for both W7
and the delayed detonation model even though we searched for the best
fit for both models varying this parameter. Although it is hard to
quantify, the fit is about the same quality as that produced by W7. The Si~II
$\lambda 6355$ line is still too fast, as as is the case for W7 the
bluer Si~II line is lost in the ``red wing'' of the redder Si~II line.
The  S~II ``W'' is reasonably well fit. The Fe~III/Mg~II/Si~II feature
just blueward of 4500~\ang doesn't match the continuum level, but the
lineshapes are about right. 
Ca~II H+K fits pretty well, but the 
Si~II/Co~II feature just to the red is not strong enough in the
emission peak. Nevertheless, all of the features observed are
reproduced in the synthetic spectrum.

\begin{figure}
\includegraphics[width=.6\textwidth,angle=90]{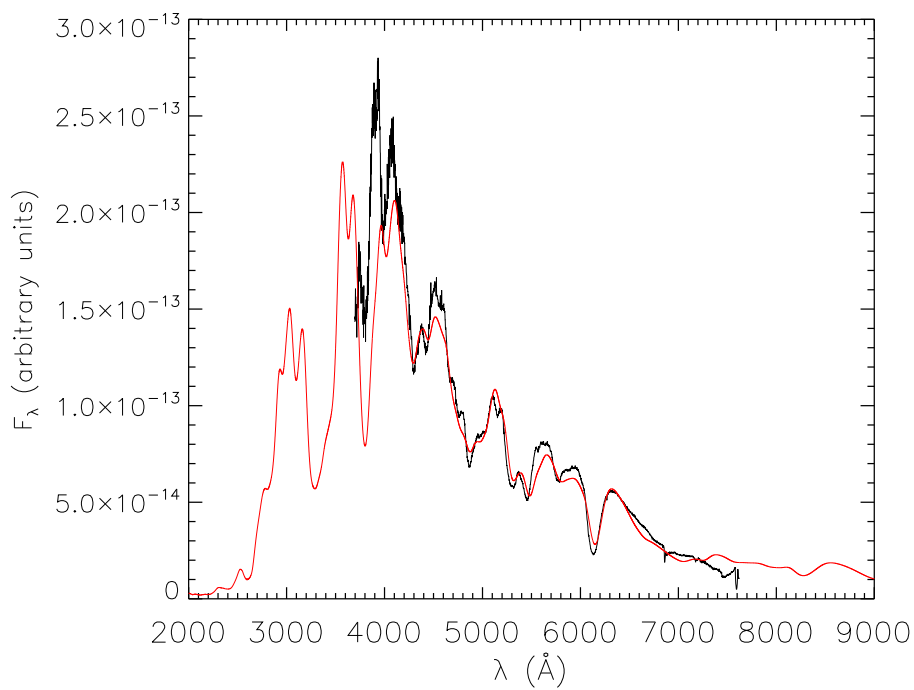}
\caption{\label{fig:94d_mar21_uniform}The
  observed spectrum of SN~1994D on March 
  21, 1994 (the time of B maximum) is compared to a synthetic spectra of a full
  NLTE parameterized model 20 days after explosion, where the
  compositions from W7 
  been artificially mixed to be uniform above the velocity point of
  8000 \kmps, the density profile has taken to be given by an
  exponential in velocity with $v_e = 1500$~\kmps, and the velocity at
  $\tstd = 1$ is 7,500\kmps. The gamma-ray deposition follows the
  density profile.}
\end{figure}

\citet{branch81b85} calculated parameterized synthetic spectra of W7
and found that mixing the compositions above 10,000~\kmps\ improved
the quality of the overall fit to SN~1981B. \citet{harkness91b} found
that unmixed models actually were better when he included line
blanketing in the UV. \citet{nug1a95} found good fits to SN~1981B and
SN~1992A with mixed W7 models. Full 3-D calculations
\citep{RH05a,gko04a,GSB05} find that deflagrations lead inevitably to
considerable mixing, so it is worth reconsidering the question of a
mixed model here. In order to study mixing we replaced W7 with a
parameterized model with a density determined by $\rho = \rho_0
e^{-v/v_e}$, where $v_e$ is the e-folding velocity and $\rho_0$ is
determined by setting the total continuum optical depth at 5000~\ang,
$\tstd= 1$, at the ``photospheric velocity'', $v_0$, and the radius at
that point $R_0 = v_0 t$, where $t$ is the time since explosion.
$v_0$ is determined by obtaining the best fit to the
lineshapes. 
As always the total bolometric luminosity in the observer's frame is
an input parameter, here it is convenient to parameterize it in terms
of the model temperature $\Tmod$. Except for the much more detailed
NLTE treatment this is 
the identical procedure to that of
\citet{nug1a95}. Figure~\ref{fig:94d_mar21_uniform} displays the
result from the best fit to the parameterized mixed calculation 20
days after explosion.  The mixed models assume  gamma-ray deposition
that follows the density profile.
While the Si~II $\lambda 6355$ line is well fit, the bluer $\lambda
5972$ line is just a bit too weak, the Si~II ``W'' is well fit although the
red edge extends a bit too far to the red.
The Si~II/Co~II feature just redward of
4000~\ang\ is too weak and Ca~II H+K is only reasonably well fit.  Since
we have altered the density structure, the compositions, and the
gamma-ray deposition of W7, it would be incorrect to ascribe the
improved fit of the Si~II $\lambda 6355$ line simply to mixed
compositions. We will discuss the line formation of Si~II $\lambda
6355$ in detail in \S~\ref{sec:discussion}

\subsection{1994D Postmax}

\begin{figure}
\includegraphics[width=.6\textwidth,angle=90]{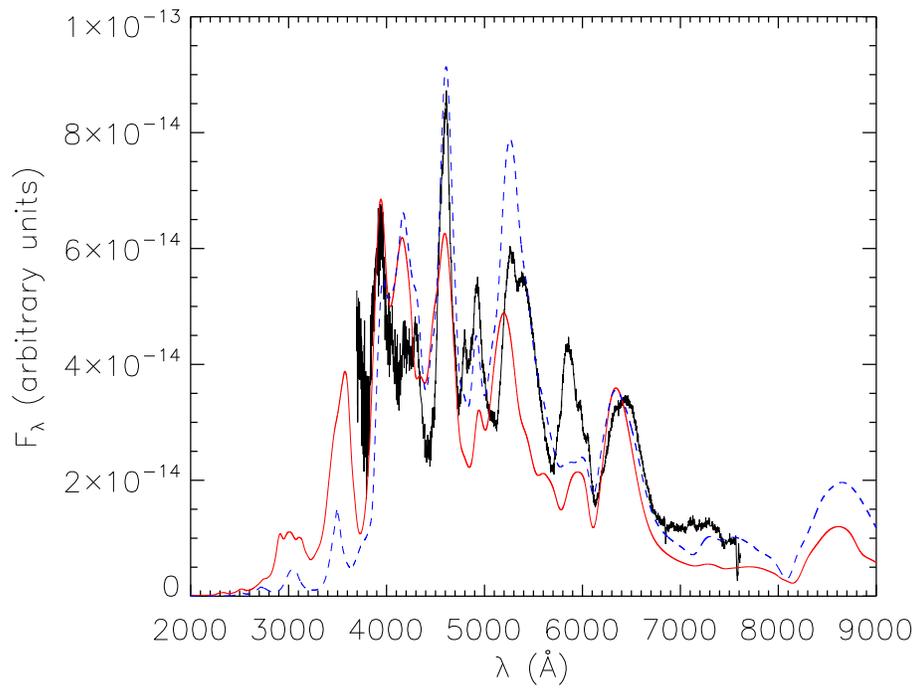}
\caption{\label{fig:pah_94d_apr05}The synthetic spectra of a full
  NLTE models 5p0z822\_16 (red line) and 5p0z822\_25 (blue line), 35
  days after explosion, are compared to the observed spectrum of
  SN~1994D obtained on Apr 5, 1994 
  (roughly 15 days after the time of B maximum). 
}
\end{figure}

Figure~\ref{fig:pah_94d_apr05} displays the synthetic spectrum of our
best fit to SN 1994D at maximum light using the 5p0z22.16/25 models of
\citet{HGFS99by02}. Apart from the glaring absence of the strong
observed Na~D line the overall fit is not bad. Here the slower
declining model 5p0z22.25 really appears to do a significantly better
job at fitting the spectrum which likely is due to the fact that the
post-maximum spectrum samples a larger amount of the ejecta, than do
spectra at maximum light.

\subsection{1999ee}

SN~1999ee is another extremely well observed SN~Ia, with spectral
coverage beginning 9 days prior to maximum light and continuing until
42 days after maximum. It was observed in the galaxy IC~5179 and was a
relatively slow decliner with $\Delta m_{15} = 0.94$
\citep{hamuy99ee99ex02}. \citet{kk04b} quote an extinction of $A_V
=0.94 \pm 0.16$ for SN~1999ee or $E(B-V) = 0.3$.

\begin{figure}
\includegraphics[width=.6\textwidth,angle=90]{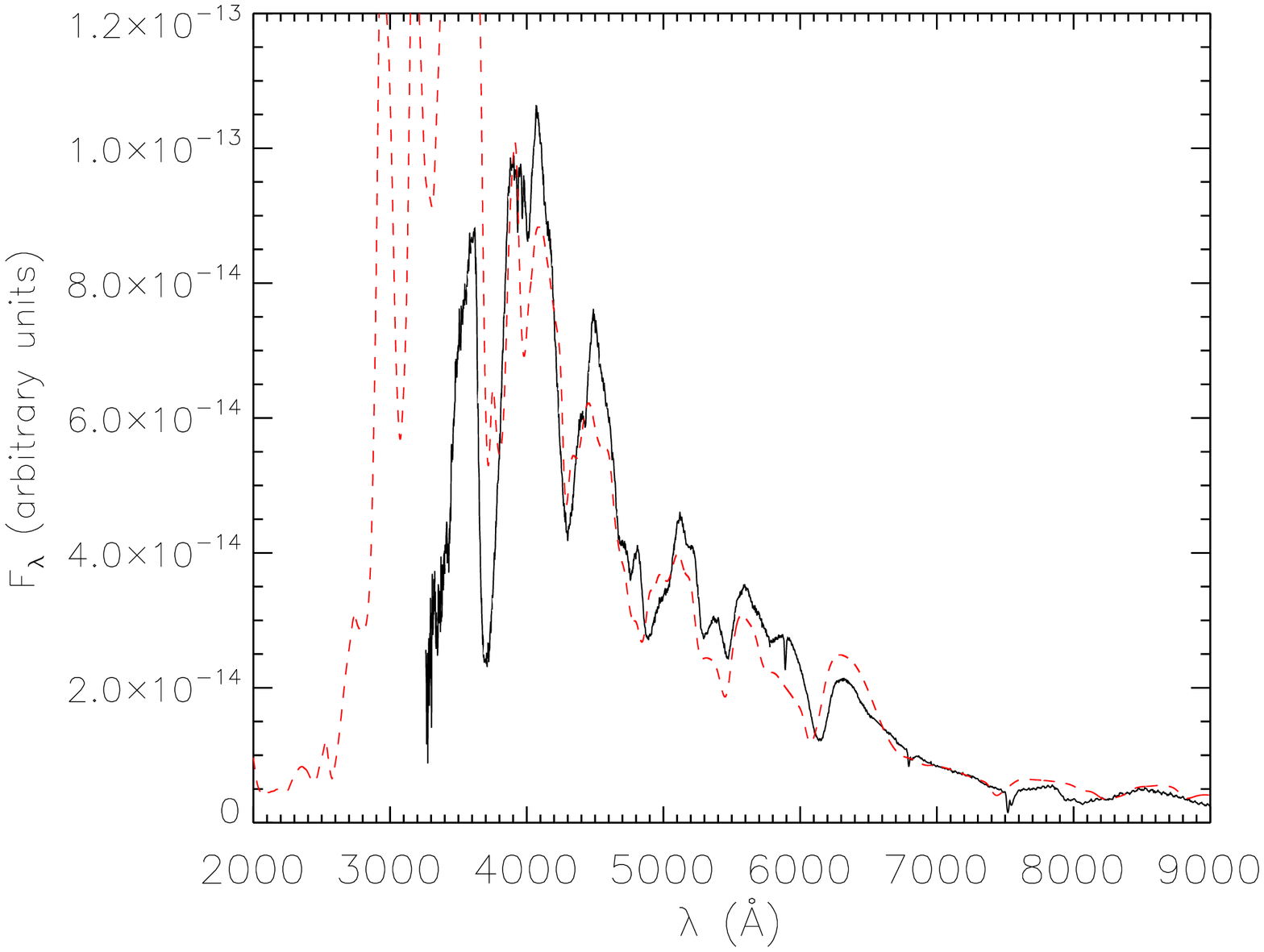}
\caption{\label{fig:w7_99ee_oct18}The synthetic spectra of a full NLTE
  model of W7, 20 days after explosion, is compared to the observed
  spectrum of SN~1999ee on October 18, 1999 (the time of B
  maximum). In this and subsequent figures on SN~1999ee the observed
  spectra have been corrected for redshift assuming a velocity of
  3498~\kmps\ \citep{hamuy99ee99ex02} and a reddening of $E(B-V) =
  0.30$ \citep{kk04b,stritz99ee99ex02}}
\end{figure}

Figure~\ref{fig:w7_99ee_oct18} compares the best fit synthetic
spectrum for the W7 model to SN 1999ee on
Oct 18, 1999 (Day -2). The overall quality of the fit is similar to that 
of SN~1994D, but with the additional wavelength coverage in the blue
part of the observed spectrum, we see that the synthetic spectrum is
too blue.

\begin{figure}
\includegraphics[width=.6\textwidth,angle=90]{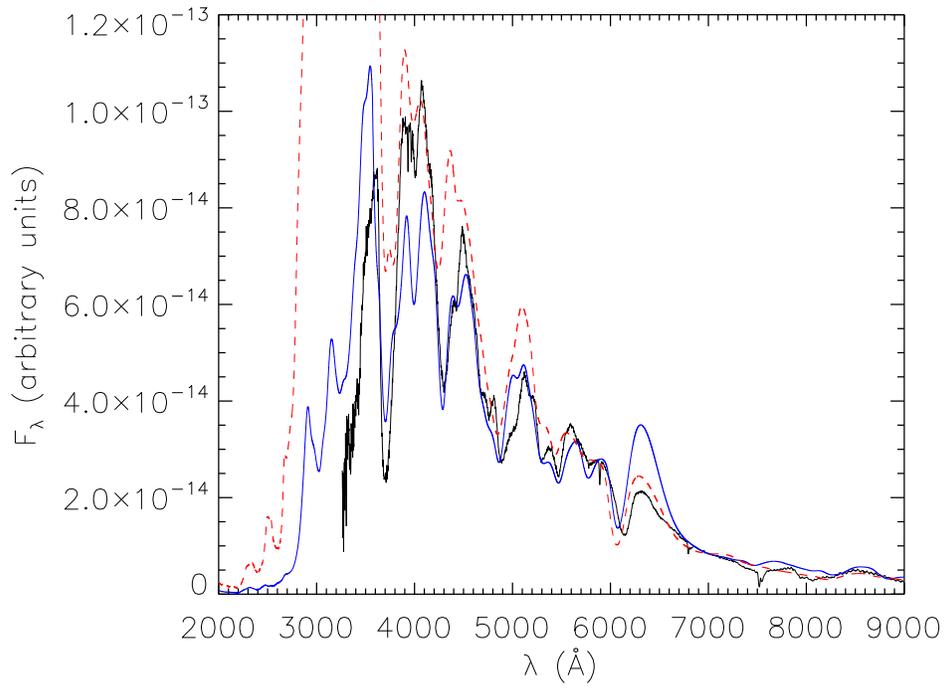}
\caption{\label{fig:pah_99ee_oct18}The synthetic spectra of a full NLTE
  models 5p0z822\_25 (blue line) and 5p0z822\_16 (red line), 20 days
  after explosion, are compared to the observed 
  spectrum of SN~1999ee on October 18, 1999 (the time of B
  maximum).}
\end{figure}

Figure~\ref{fig:pah_99ee_oct18} compares the best fit synthetic
spectrum for the two delayed-detonation models to SN 1999ee on Oct 18,
1999. Since model 5p0z22.25 has a $\Delta m_{15}$ closer to that of
SN~1999ee (although it still declines a little too rapidly) should be
a better fit to the observed spectrum.  However, the fast decliner
5p0z22.16 fits much better (as was the case for SN~1994D, since the
spectra at maximum are quite similar). This suggests that for spectra
formation near maximum light the amount of nickel produced is not so
important (if there is enough energy to reach the luminosity
required).
For the slow decliner 5p0z22.25, the lineshapes are
overall not bad, but the continuum is too blue, 
particularly to the blue of Ca~II H+K. However, the Si~II $\lambda
5972$ line is stronger for the fast decliner, which is in the same
sense as observed empirically.

\begin{figure}
\includegraphics[width=.6\textwidth,angle=90]{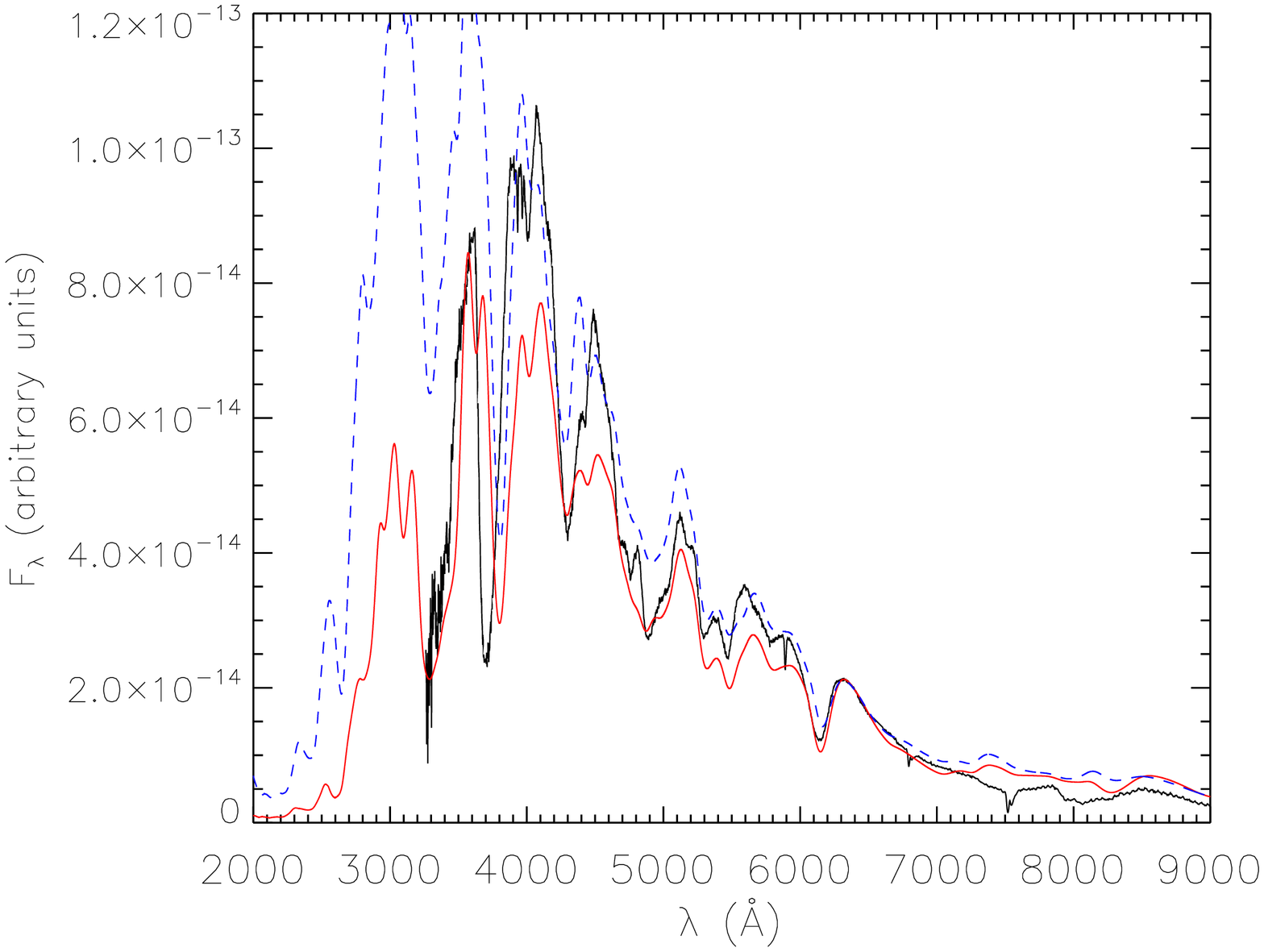}
\caption{\label{fig:w7_99ee_oct18_day20_uniform}
The observed spectrum of
  SN~1999ee on October 
  18, 1999 (the time of B maximum) compared with a synthetic spectra of a full
  NLTE model 20 days after explosion, where the W7 compositions have
  been artificially mixed to be uniform above the velocity point of
  8000 \kmps,  the density profile has taken to be given by an
  exponential in velocity with $v_e = 1500$~\kmps,  and the velocity at
  $\tstd = 1$ is 7,500~\kmps. The gamma-ray deposition follows the
  density profile. The red line has $\Tmod=10,500$~K and the blue line
$\Tmod=11,500$~K}
\end{figure}

Figure~\ref{fig:w7_99ee_oct18_day20_uniform} compares the best fit synthetic
spectrum for the parameterized mixed model on day 20 to SN 1999ee on
Oct 18, 1999 for two choices of $\Tmod$, 10,500~K and 11,500~K. The
other parameters are the same as those for the 1994D 
uniform-composition model. The fit is not bad, the Ca~II H+K is too
slow, but the 
overall continuum shape is reasonably well reproduced, for the
10,500~K model. 

\begin{figure}
\includegraphics[width=.6\textwidth,angle=90]{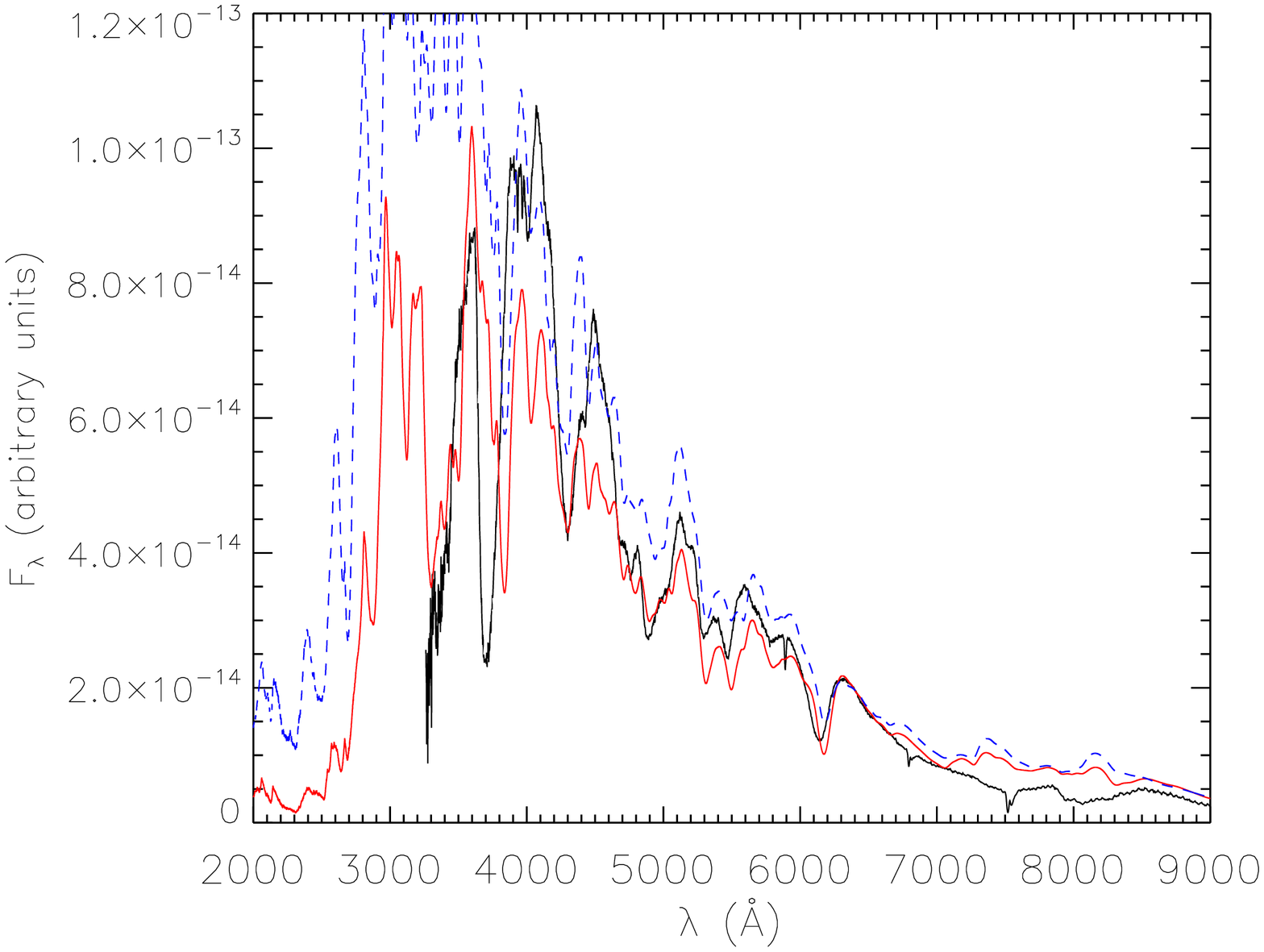}
\caption{\label{fig:w7_99ee_oct18_day15_uniform}The observed spectrum of
  SN~1999ee on October 
  18, 1999 (the time of B maximum) compared with a synthetic spectra of a full
  NLTE model 15 days after explosion, where the W7 compositions have
  been artificially mixed to be uniform above the velocity point of
  8000 \kmps,  the density profile has taken to be given by an
  exponential in velocity with $v_e = 900$~\kmps. There is no
  gamma-ray deposition. The red line has $\Tmod=11,000$~K and the blue line
$\Tmod=12,000$~K}
\end{figure}

It is possible that slow decliners have somewhat higher densities
(which leads to higher opacities and thus longer risetimes). To test
this we calculated a uniform model but at an earlier time. The
velocity at the $\tstd =1$ was the same, but the e-folding velocity
was smaller leading to a steeper density profile and hence higher
density at $\tstd = 1$. The result is shown in
Figure~\ref{fig:w7_99ee_oct18_day15_uniform}.  Although some of the
lines are too narrow the fit is not bad, and the lines are forming at
the right velocities. A model with an increased density can produce a
reasonable fit. The narrowness of the lines in our synthetic spectrum
is due to the choice of a steeper density profile, the lines could be
made wider with a shallower density profile.

\subsection{1992A}

\begin{figure}
\includegraphics[width=.6\textwidth,angle=90]{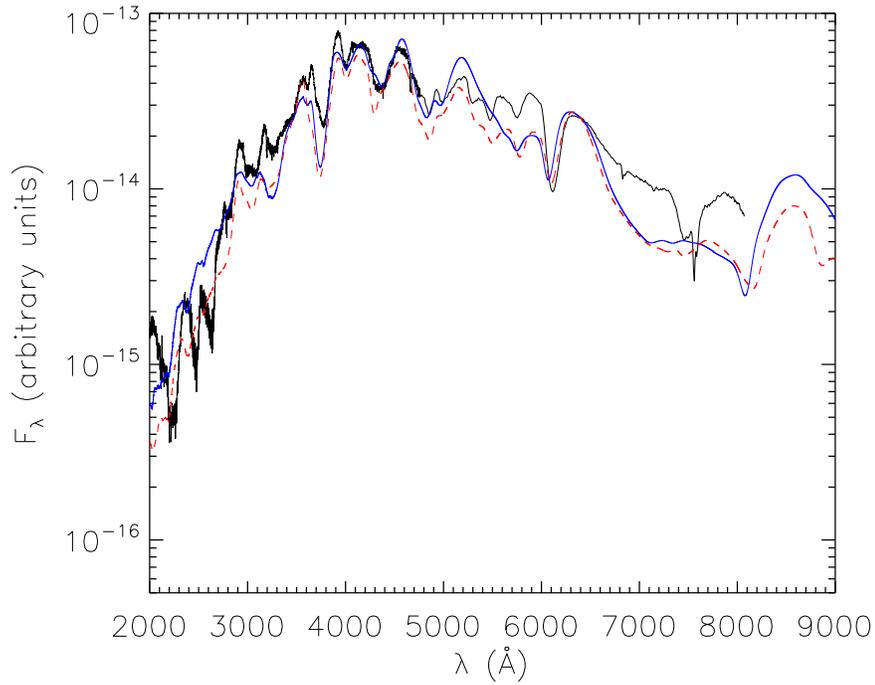}
\caption{\label{fig:pah_92a_jan24}The synthetic spectra of a full
  NLTE models 5p0z822\_25 (blue line) and 5p0z822\_16 (red line), 25
  days after explosion, are compared to the observed spectrum of
  SN~1992A obtained on Jan 25, 1992 (optical) and Jan 24, 1992 (HST)
  (roughly 5 days after the time of B maximum). The observed spectra have
  been corrected for redshift assuming a velocity of 1845~\kmps\ and
  no reddening correction has been applied.
}
\end{figure}

The availability of an HST spectrum of SN~1992A provides a good
opportunity to look at a broader wavelength range.
Figure~\ref{fig:pah_92a_jan24} compares the best fit synthetic
spectrum for the two delayed-detonation models to SN 1992A on
Jan 24, 1992 (HST) and Jan 25, 1992 (optical), roughly 5 days past
maximum light. The fit is adequate, the overall shape of the spectrum
is well reproduced, but most of the lines and continua do not fit well.

\section{Discussion\label{sec:discussion}}

\begin{figure}
\includegraphics[width=.6\textwidth,angle=90]{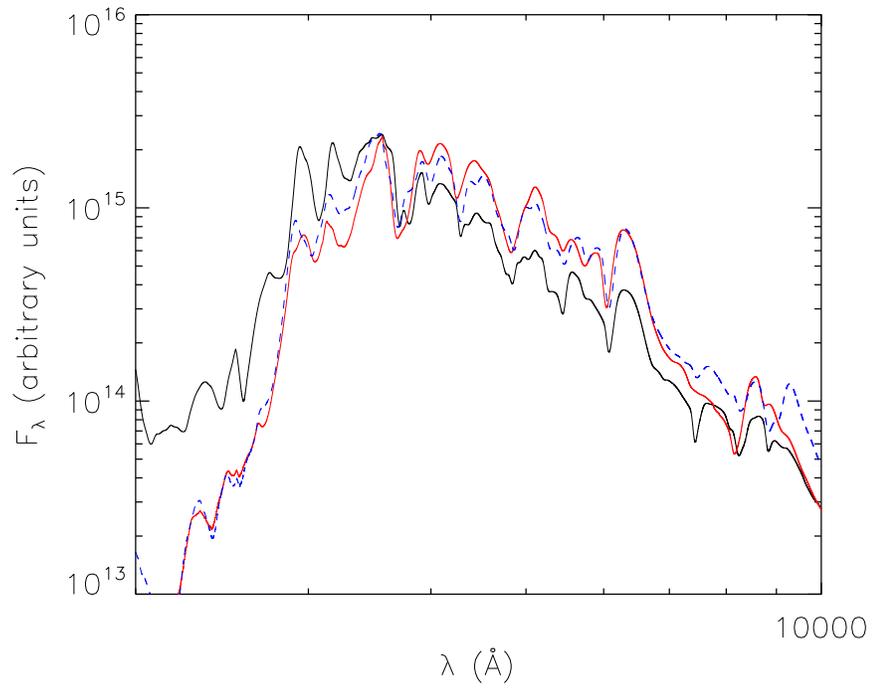}
\caption{\label{fig:w7_v_pah_day20}The synthetic spectra of a full
  NLTE models of W7 (black line), 5p0z822\_16 (red line) and 5p0z822\_25
  (blue line), 20 
  days after explosion, are compared to each other.
}
\end{figure}

As we have showed above, the optical spectra
of W7 are similar in many respects to those of the delayed-detonation
models, nevertheless the distribution of the energy over a wider
wavelength region is significantly different.
Figure~\ref{fig:w7_v_pah_day20} compares the day 20 spectra of W7,
5p0z22.16 and 5p0z22.25 where all models have the same bolometric
magnitude. Clearly the W7 spectrum has a much smaller UV 
deficit than do the delayed-detonation models. This is illustrated in
Table~\ref{tab:mags} where the absolute magnitudes in a wide range of
photometric bands is listed. Clearly W7 is much bluer than the delayed
detonation models even though we have kept $M_{bol}$ fixed. All of the
models are quite ``blue'' in $U-B$, but that is an intrinsic feature
of SN~1994D.

\begin{deluxetable}{llll}
\tablecolumns{4}
\tablewidth{0pc}
\tablecaption{\label{tab:mags}}
\tablehead{
\colhead{Absolute Magnitude}    &   \colhead{W7}   &
\colhead{5p0z22.16} & \colhead{5p0z22.25}}
\startdata
$M_{bol}$&-19.1& -19.1 & -19.1\\
$M_U$& -20.10 & -19.92 & -19.82\\
$M_B$& -19.01 & -19.29 & -19.45\\
$M_V$& -18.72 & -19.22 & -19.32\\
$M_R$& -18.66 & -19.22 & -19.18\\
$M_I$& -18.26 & -18.66 & -18.44\\
$M_Z$& -18.27 & -18.68 & -18.54\\
$M_J$& -17.55 & -18.21 & -17.60\\
$M_H$& -17.41 & -17.77 & -17.51\\
$M_K$& -17.24 & -16.79 & -16.79\\
\enddata
\end{deluxetable}

\begin{figure}
\includegraphics[width=.6\textwidth,angle=90]{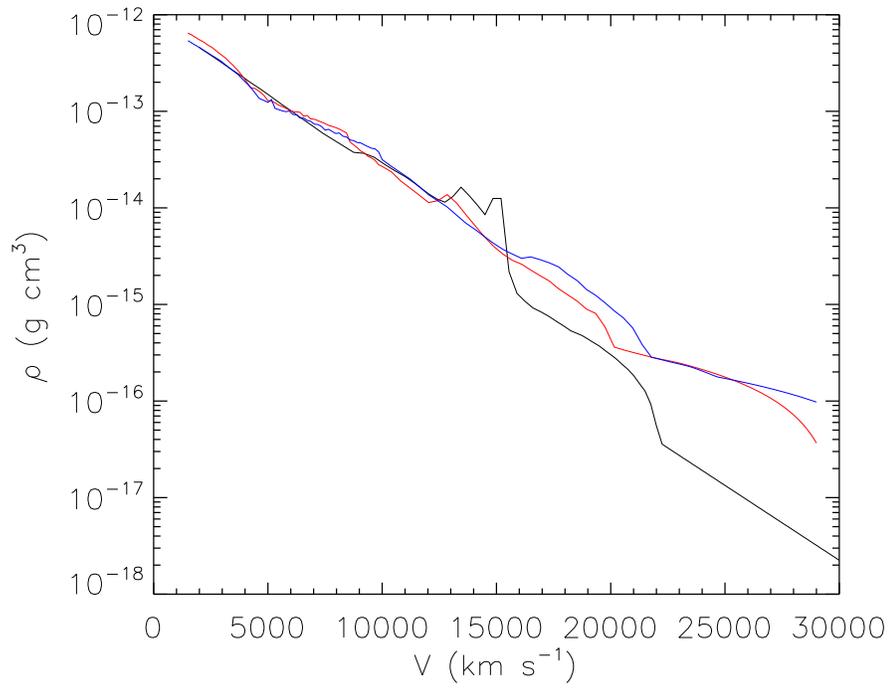}
\caption{\label{fig:rho_w7_v_pah_day20}The density structure of W7 (black
  line), 5p0z822\_16 (red line), and 5p0z822\_25 (blue line), 20 days after explosion, are
  compared to each other.}
\end{figure}

Figure~\ref{fig:rho_w7_v_pah_day20} compares the density profiles at
day 20 of W7 to 
5p0z22.16 and 5p0z22.25. The densities in the inner parts are nearly
identical for all the models, W7 exhibits a characteristic ``bump''
where the deflagration dies out around 15,000 \kmps, and the delayed
detonation models have significantly higher densities than W7 in the
outer parts.

\begin{figure}
\includegraphics[width=.6\textwidth,angle=90]{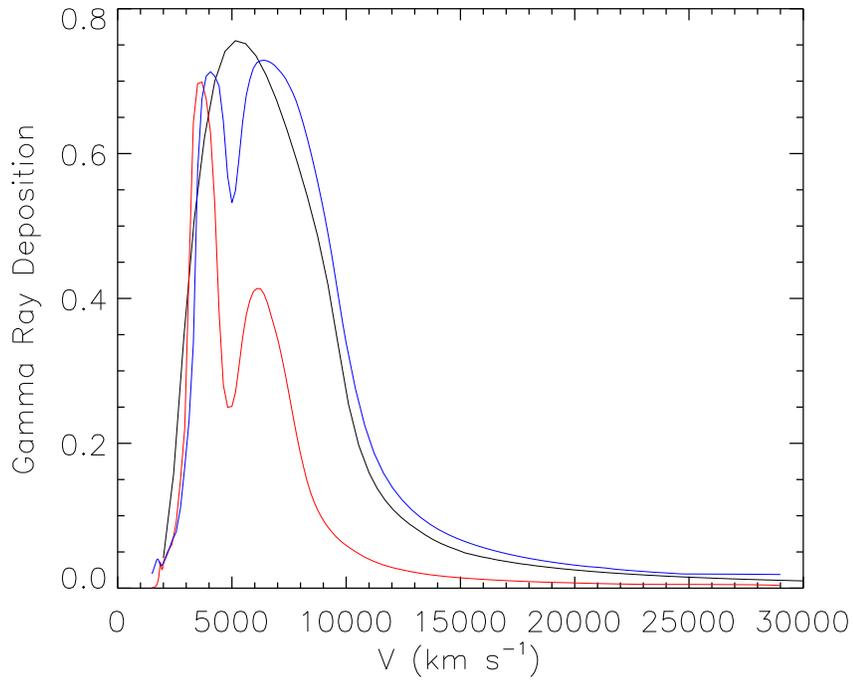}
\caption{\label{fig:gamdep_w7_v_pah_day20}The gamma-ray deposition of W7 (black
  line), 5p0z822\_16 (red line), and 5p0z822\_25 (blue line), 20 days
  after explosion, are 
  compared to each other.}
\end{figure}

Figure~\ref{fig:gamdep_w7_v_pah_day20} compares the gamma-ray
deposition function at day 20 of W7 to 5p0z22.16 and 5p0z22.25, except
for the pronounced dip at the deflagration to detonation transition
the shape of the deposition function is very similar to that of W7,
particularly for model 5p0z22.25, which is about the same nickel mass
as W7. In fact it is interesting to note that we obtained our best fit
to SN 1994D at maximum light with model 5p0z22.16 which has only about
0.27~\msol\ of \nni, not enough to produce the observed brightness of
normal SNe~Ia.

\begin{figure}
\includegraphics[width=.6\textwidth,angle=90]{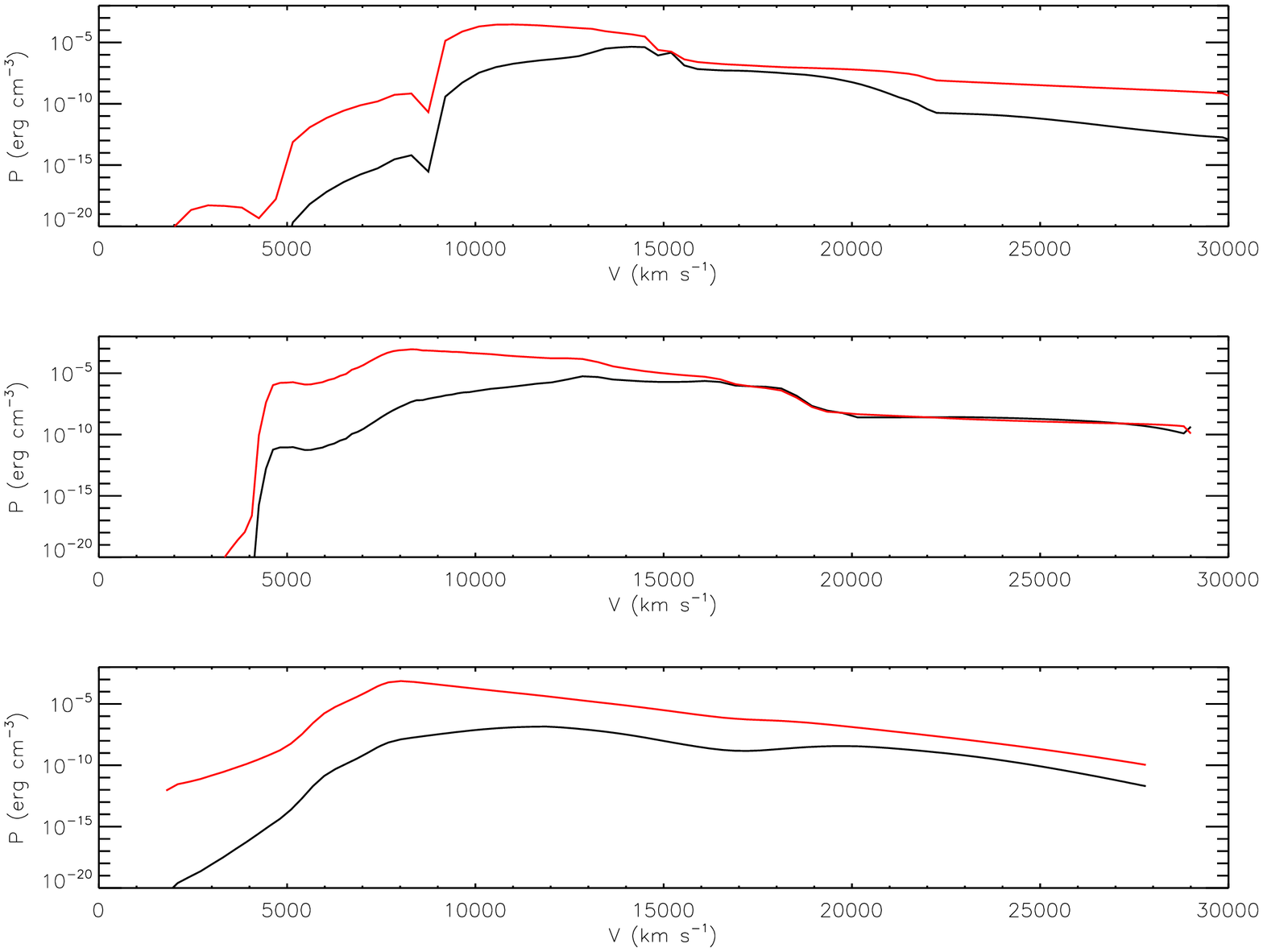}
\caption{\label{fig:ppres_si}The partial pressures of 
  Si~II (black line) and Si~III (red line) for W7 (top panel), the
  delayed detonation model 5p0z822\_16 (middle panel), and the mixed
  model (bottom panel). Each model corresponds to the best fit for
  that particular model to SN~1994D.}
\end{figure}

\begin{figure}
\includegraphics[width=.6\textwidth,angle=90]{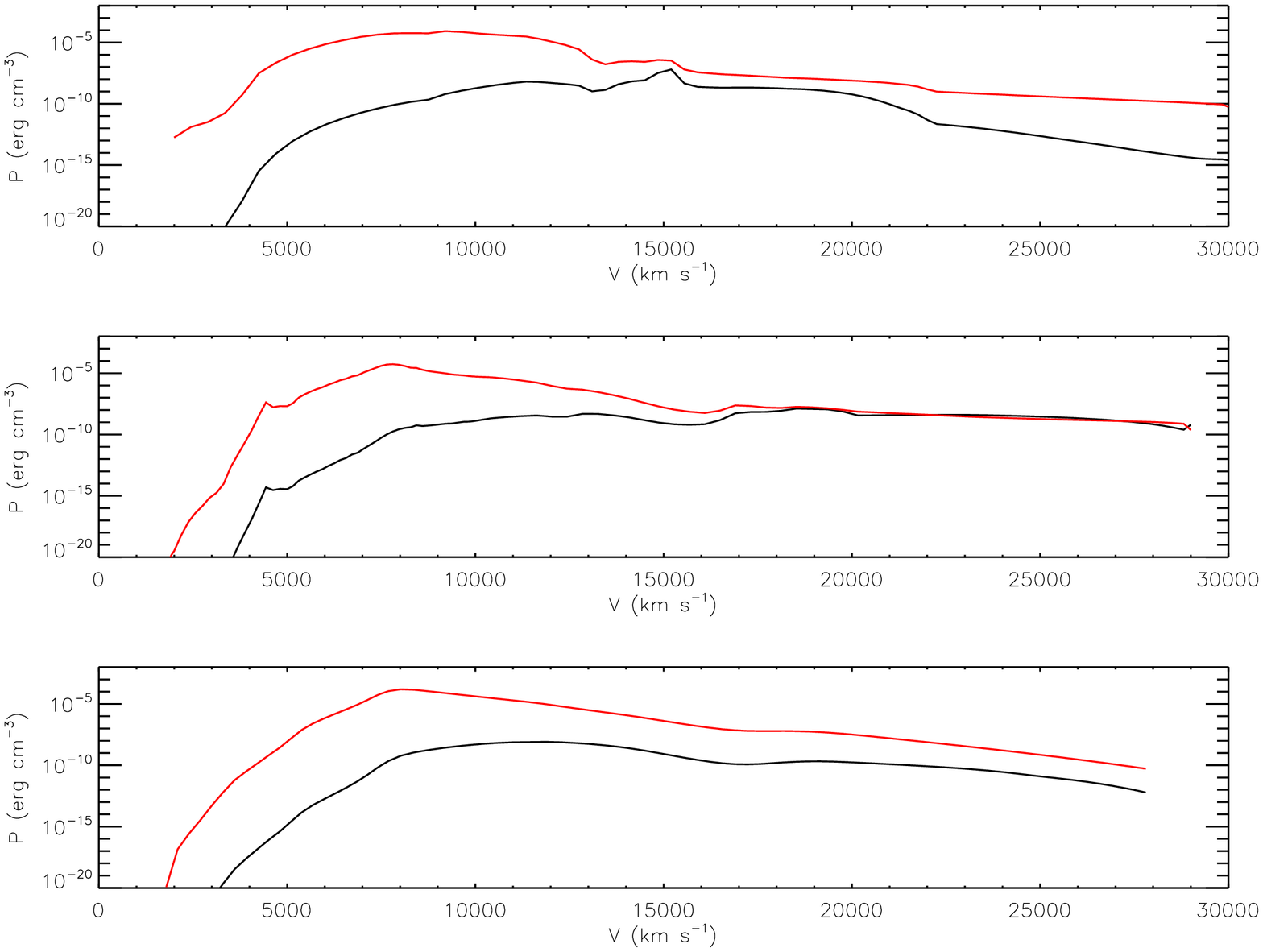}
\caption{\label{fig:ppres_fe}The partial pressures of 
  Fe~II (black line) and Fe~III (red line) for W7 (top panel), the
  delayed detonation model 5p0z822\_16 (middle panel), and the mixed
  model (bottom panel). Each model corresponds to the best fit for
  that particular model to SN~1994D.}
\end{figure}

It is not easy to identify exactly why the Si~II line absorption
trough is too blue in W7 (and to a lesser extent in the delayed
detonation models). It is tempting to point to the steep outer density
gradient in W7 as the primary difference, however the uniform-composition
models have an even steeper density gradient in the outer parts than
does W7 and they in fact, fit the Si II feature best. The 
iron mass fraction is higher in the uniform models than in either W7
or in the delayed detonation models. Figs.~\ref{fig:ppres_si} and
\ref{fig:ppres_fe} display the partial pressures of Si~II--III and
Fe~II--III, respectively. The obvious difference between W7 and the
other two models is that Fe~III/Fe~II and Si~III/Si~II strongly
increases with velocity in W7, whereas it is nearly flat in both the
delayed detonation model and in the uniform composition
model. \citet{stehle02bo05} found good fits to the 
Si~II $\lambda 6355$ feature of SN 2002bo near maximum light, using
the density profile of W7, but varying the compositions. They also
found that they needed more iron in the outer parts of the atmosphere
than W7 has. Interestingly, even with a large number of parameters,
they weren't able to well reproduce the sulfur ``W'' feature.
\citet{stehle02bo05} use a parameterized Schuster-Schwarzschild
model and the 
Sobolev approximation and thus do not solve the full NLTE radiative
transfer problem. However, they have fit the composition structure of
by running a very large set of models (their code, with its simpler
assumptions requires significantly less computer resources than do our
models). Thus, given the fact that that they have altered the
abundances in the model to get the best fit, they still do not
reproduce the sulfur W, which leads us to suspect that it is the
density structure (in a complicated combination with the abundances)
that produces this feature and that is just what we have attempted to
describe in this work.

Both the blue color of W7 and the fact
that higher ionization stages increase with velocity leads us to
conclude that W7 is ``over-ionized''. This is not due to increased
gamma-ray deposition as is clear from
Fig.~\ref{fig:gamdep_w7_v_pah_day20}, but rather due to the steep
density fall-off, or more correctly the steep fall-off of iron in the
outer parts of W7. This steep fall-off in the iron abundance significantly
reduces the UV deficit and thus there are more ionizing photons
available to ionize the outer parts of the atmosphere.

It is clear (see for example
Figure~\ref{fig:w7_99ee_oct18_day20_uniform}) that the blue part of the 
spectrum is extremely sensitive 
to the outer boundary condition (the total bolometric luminosity in
the observer's frame), thus the failure to fit the Si II and S II
features seems to be generic to the models and not just an artifact of
our choice of $M_{bol}$.

\citet{bran94d05} found that Si~III was required in a velocity range
from about 10,000--17,000 \kmps\ and that Si~II was required from
about 8,000--22,000 \kmps. Figure~\ref{fig:ppres_si} shows that all
the models have significant Si~II-III all the way out to the
outermost layers of the atmosphere.

That Si~II $\lambda 6355$ forms at high velocity in W7 even though the
ratio of Si~III/Si~II is increasing is not surprising. The lower level
is at 8.2eV, so some ionization actually will strengthen this line.
Also, as we discuss in detail in a future publication (S.~Bongard
\etal, in preparation) the actual line profile of the Si~II $\lambda
6355$ and Si~II $\lambda 5972$ lines is more likely due to complex
radiative transfer effects involving Fe~II and Fe~III lines.

\citet{quimby05cg06} studied the early spectra of SN~2005cg and
concluded that the triangular shaped \SiIIred\ feature was a unique
prediction of the delayed-detonation models. However,
Figure~\ref{fig:w7_94d_mar21} clearly shows that W7 produces just such
a line as late as maximum light. A direct confrontation of the spectra
of W7 and 
the delayed-detonation models with the spectra of SN~2005cg will be
the subject of future work.

\section{Conclusions}

Overall both the highly parameterized deflagration model W7 and the
delayed detonation models that we have studied provide about equally
good fits to spectra of normal SNe~Ia near maximum light. The quality
of the fit to near maximum light spectra is not strongly dependent on
the distributiion of nickel (or gamma-ray deposition) for the delayed
detonation 
models (although we have forced the models to have the ``right''
luminosity, which is not self-consistent). W7 produces spectra that
are both too blue, and where the Si~II $\lambda 6355$ line (the
defining feature of SNe~Ia) is significantly too fast. This appears to
be due to the rapid fall-off in the amount of iron in the outer layers
of W7 and can be remedied either by flattening the density profile, or
by increasing the relative fraction of iron. Our best fit model for
the \SiIIred\ feature near maximum had uniform compositions and a
steeper overall density decrease than W7, which shows that the blue
wing of the Si~II feature found in W7 is due to material at high
velocity and can be supressed simply by removing that
material. Clearly successful hydrodynamical models of SNe~Ia require
stratified compositions, nevertheless somewhat larger amounts of iron
(whether it be primordial or produced in the ejecta) seem to help. The
delayed detonation models did a better job at reproducing the observed
position of the Si~II $\lambda 6355$ line and simply have a flatter
density profile, so that may be an indication that flatter density
profiles, as well as more iron in the outer layers is required.

\begin{acknowledgments}
We thank the anonymous referee for comments that improved the
presentation.
This work was supported in part  by NASA grants
NAG5-3505 and NAG5-12127, and NSF grants AST-0204771 and AST-0307323, 
PHH was 
supported in part by the P\^ole Scientifique de Mod\'elisation
Num\'erique at ENS-Lyon. EB thanks the Laboratoire de Physique
Nucl\'eaire et de Haute Energies, CNRS-IN2P3, 
University of Paris VII, for a Professeur Invit\'e when much of this
work was done.
This research used resources
of: the San Diego Supercomputer Center (SDSC), supported by the NSF;
the National Energy Research Scientific Computing Center (NERSC),
which is supported by the Office of Science of the U.S.  Department of
Energy under Contract No. DE-AC03-76SF00098; and the
H\"ochstleistungs Rechenzentrum Nord (HLRN).  We thank all these
institutions for a generous allocation of computer time.
\end{acknowledgments}


\end{document}